\documentclass[fleqn,10pt]{wlscirep}
\usepackage[utf8]{inputenc}
\usepackage[T1]{fontenc}

\title{Temporal visitation patterns of points of interest in cities on a planetary scale: a network science and machine learning approach}

\author[1,+,*]{Francisco Betancourt}
\author[1,+]{Alejandro P. Riascos}
\author[1,2,+]{José L. Mateos}
\affil[1]{Instituto de Física, Universidad Nacional Autónoma de México, Ciudad Universitaria, 04510 Ciudad de México, México.}
\affil[2]{Centro de Ciencias de la Complejidad, Universidad Nacional Autónoma de México, 04510 Ciudad de México, México.}

\affil[*]{email: cfrancisco@ciencias.unam.mx}

\affil[+]{These authors contributed equally to this work}

\begin{abstract}
We aim to study the temporal patterns of activity in points of interest of cities around the world. In order to do so, we use the data provided by the online location-based social network Foursquare, where users make check-ins that indicate points of interest in the city. The data set comprises more than 90 million check-ins in 632 cities of 87 countries in 5 continents. We analyzed more than 11 million points of interest including all sorts of places: airports, restaurants, parks, hospitals, and many others. With this information, we obtained spatial and temporal patterns of activities for each city. We quantify similarities and differences of these patterns for all the cities involved and construct a network connecting pairs of cities. The links of this network indicate the similarity of temporal visitation patterns of points of interest between cities and is quantified with the Kullback-Leibler divergence between two distributions. Then, we obtained the community structure of this network and the geographic distribution of these communities worldwide. For comparison, we also use a Machine Learning algorithm - unsupervised agglomerative clustering - to obtain clusters or communities of cities with similar patterns. The main result is that both approaches give the same classification of five communities belonging to five different continents worldwide. This suggests that temporal patterns of activity can be universal, with some geographical, historical, and cultural variations, on a planetary scale.      

\end{abstract}
\begin{document}

\flushbottom
\maketitle
\section*{Introduction}
In recent years, cities have become a topic of considerable scientific interest\cite{batty2013new, barthelemy2016structure, bettencourt2021introduction}. In the last decade, the development of technologies applied to inform the activities of humans has made available an unprecedented amount of data associated with the digital trace of humans in cities\cite{UrbanInformaticsBook,rybski2022cities}. For instance: the use of credit card transactions\cite{sobolevsky2016cities}, the use of digital devices to access transportation services, like taxis\cite{riascos2020networks}, buses and subway\cite{ChapterMateosBook}, bicycle \cite{loaiza2019human}, and telecommunications networks like cell phones\cite{gonzalez2008understanding, song2010modelling, louail2014mobile, louail2015uncovering, ccolak2016understanding, alessandretti2018evidence, song2010limits} and GPS devices\cite{alessandretti2020scales, KaskiReview}. 
\\[2mm]
A very important source of information for the study of human behavior in cities are Location-Based Social Networks (LBSN)\cite{wei2022survey}. In these, people share information about the places they visit that may be of interest to other people in this social network. These venues, also known as Points of Interest (POIs), correspond to places within the city that can be characterized by features, such as restaurants, bars, gyms, hospitals, museums, parks and so on. Besides the type of feature, LBSN identify, for each POI, the spatial location (coordinates or addresses) and the time of visit (date and hour); these visits are recorded as check-ins using, typically, a mobile-phone application. Therefore, we end up with a data set with detailed information of the type of place and the exact location, together with the day of the week, and time of the visit. LBSN provide valuable information on the interaction of people among themselves, as well as the physical places where these interactions occur. This interplay has been explored in studies of urban mobility\cite{chen2022contrasting}, human behavior\cite{lenormand2015human, yang2020location2}, social interactions\cite{yang2019revisiting} and encounter networks\cite{riascos2017emergence}.
\\[2mm]
In this study, we use data from one of the more relevant location-based social networks: Foursquare\cite{Foursquare}. This LBSN has been used previously in several studies\cite{yang2019revisiting, riascos2017emergence, noulas2012tale, Yang15, yang2020lbsn2vec++, yang2020location, gallotti2021unraveling, noulas2015topological, noulas2011empirical, cornacchia2020modelling, d2018predicting}, of human mobility and social relationships, encounter networks due to human mobility, spatial and temporal activity patterns, among others. A more detailed description of Foursquare will be given in the next sections. In this paper, we will focus mainly on the temporal patterns of human activities in cities throughout the world.  This temporal signature of the vastness of human dynamics can be captured through time series, using different sources such as email records\cite{barabasi2005origin}, mobile phone calls\cite{saramaki2015seconds}, data sets from the public sector\cite{prieto2021heartbeat} and the analysis of the frequency of check-ins during the visitation of POIs in cities\cite{sparks2020global}. In this work, we will study the temporal patterns that emerge due to the frequent visitation of points of interest in different cities around the world. The data set we analyzed contains more than 11 million POIs of 632 cities of 87 countries located in 5 continents. We will focus mainly on the temporal patterns of activities on a weekly basis, that distinguish weekdays and weekend.  Using network science\cite{BarabasiBookNS}, we generate an undirected weighted network connecting pairs of cities. The links between each pair of cities have a weight that quantifies the similarity of the temporal patterns of visitation of POIs. To measure the similarity between the two distributions of temporal patterns, we used the relative entropy (Kullback-Leibler divergence)\cite{CoverThomasBook} between the corresponding distributions. Once we have our network of cities, we obtained the community structure of this network using a well-known method of community detection: the Louvain algorithm. Then we locate these communities of cities geographically and notice a strong correlation between geography and temporal patterns of activities. On the other hand, besides the network science approach, we use a Machine Learning approach to classify the clusters in the data set of temporal distributions of activity. In particular, we used the unsupervised agglomerative hierarchical clustering algorithm\cite{MLBook1} to obtain clusters or communities of cities that have similar time distributions of check-ins in Foursquare. Both approaches, network science and machine learning, give a very similar classification of five communities that distribute geographically in five different continents throughout the world.
\\[2mm]
{The dataset used comprises over 90 million records made in more than 11 million POIs. Each record contains information about the place and time at which a person visited a POI. Additionally, using another dataset constructed from public information on Foursquare, it is possible to know the precise coordinates of venues and the type of place based on different categories used by this social network to classify sites. Thus, this is a database that provides spatial and temporal information on 2,733,324 individuals around the world, as well as information on the type of activities they perform. In this manner, we obtained spatial and temporal patterns of activity in many cities. In this work, we will focus mainly on the temporal patterns of activities on a weekly basis.}
\\[2mm]
Finally, it is worth mentioning part of the motivation that led to this research. The origins of the present paper can be traced back to our interest in the interplay between human mobility in cities and the encounter (or contact) networks that emerge. In a paper published in 2017\cite{riascos2017emergence} by two of the authors, we explore precisely this emergence by analyzing data from Foursquare in two cities: New York City and Tokyo. We had the spatial and temporal data corresponding to the check-ins of many users visiting POIs in both cities. Thus, we explore the co-occurrences of two users in the same place at the same time in order to obtain the encounter or contact network. The motivation was, among other things, to study the propagation of viruses during an epidemic. By the way, this study was done and published previously to the current COVID-19 pandemic, where these contact networks were important to track the onset of the pandemic at the beginning of 2020. Years later, we keep studying the rich data set of Foursquare, but, instead of two cities, we explore more than 600 cities around the globe. In particular, the points of interest in cities are precisely the places where many people congregate and therefore can lead to an explosion of ideas and innovation, but at the same time can lead to the origin of epidemics of infectious diseases. That is part of our motivation to study in more detail the mobility and the temporal patterns of activity of POIs and to make a quantitative comparison of these patterns using metrics such as the Kullback-Leibler divergence. With this metric, we constructed a network of cities on a planetary scale and obtained the community structure of this network.         
\\[2mm]
The paper is structured as follows: After the Introduction, we show first the Results with all the details in separate sections, then a Discussion and finally the Methods we employ.          
\section*{Results}
\subsection*{Check-ins and  points of interest in Foursquare}
\begin{figure}
    \centering
    \includegraphics[width=0.95\linewidth]{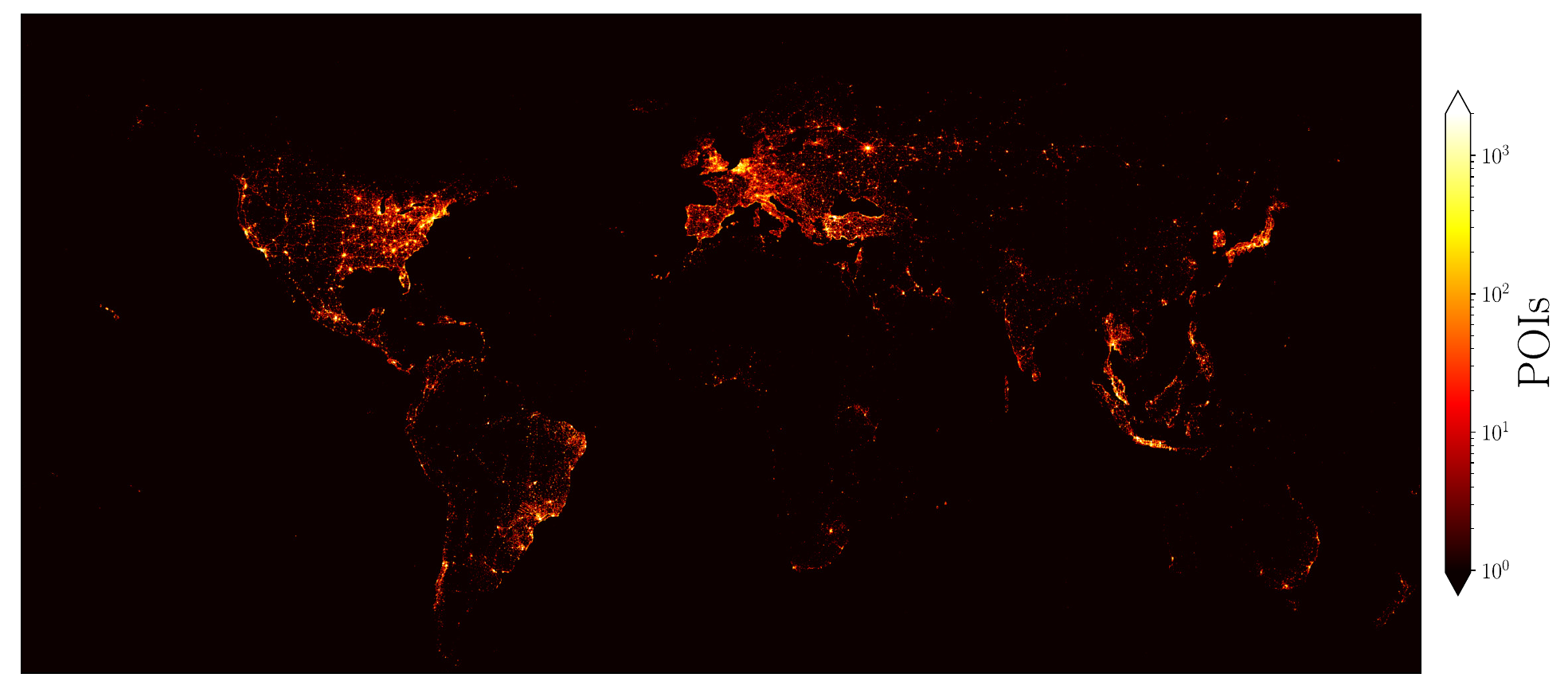}
    \caption{{\bf Points of interest worldwide from human activity in Foursquare.} This analysis includes 11,179,790 points of interest (POIs). The counts codified in the colorbar show the number of POIs in each rectangle of a grid from $-180^\circ$ to $180^\circ$ in the longitude and from $-56^\circ$ to $85^\circ$ in the latitude (the dimensions of the grid are $3,600\times 1,410$ squares defined by sides with $0.1^\circ$). A logarithmic scale was used to show the non-null frequencies of POIs found in the squares; the zones with null counts of POIs are depicted in black. This representation only considers geographical information of POIs, no map was used for this analysis.  This figure was created using python 3.8 and the matplotlib (3.5.0) package (\url{https://matplotlib.org}). }
    \label{Fig_1}
\end{figure}
In this work, we use Foursquare check-ins as a proxy of human activity in cities. Check-ins are spatio-temporal interactions between users and points of interest (POIs). They provide vast information about people's interests, site characteristics, and behavioral patterns in cities, among many others. This is why Foursquare metadata and location-based social networks in general, can provide useful information for studies of mobility, infrastructure, human behavior, and public policy, just to mention a few examples.
\\[2mm]
We study a large-scale and long-term Foursquare data set collected by Yang et al. \cite{yang2019revisiting,yang2020lbsn2vec++} that is publicly available \cite{Yang-HomePage}. The data set contains 90,048,627 check-ins made by 2,733,324 users from April 2012 to January 2014. The authors collected active check-ins from Twitter by searching a hashtag generated automatically for users that linked their Foursquare activity with that social network. Then, with the set of check-ins where venues occurred, they collected the POIs description; this information is available to the public from the Foursquare platform. The records include detailed information about the POIs, their coordinates, and the exact time of each user check-in \cite{yang2019revisiting,yang2020lbsn2vec++}. In this manner, the database provides the space-temporal activity obtained from the sequence of successive check-ins for each user. Furthermore, the registers include a short description that labels each  POI with specific keywords  (for example ``restaurant'', ``metro'', ``university'') and a social network of users defined via the mutual following between Twitter accounts. We did not incorporate this information in our study (see the methods section for a detailed description of the data).
\\[2mm]
The complete dataset includes information of Foursquare users in every country in the world. From the 253 country codes, check-ins on countries with 5000 or more POIs  represent 98.92\% of data. Almost all of the data is concentrated in a third of the countries. To visualize the geographical distribution of the POIs worldwide, we grouped them according to their coordinates. We generate a two-dimensional histogram dividing the latitude and longitude using square bins with side $0.1^\circ$ and counting the number of POIs in each coordinate square. This allowed us to identify regions with non-null POIs and those with the highest concentration. In Fig. \ref{Fig_1} we depict the spatial distribution obtained; it consists of $3600 \times 1410$ squares covering latitudes from $-56.6^\circ$ and $84.6^\circ$, and longitudes from $-180^\circ$ to $180^\circ$; latitudes excluded are due to the lack of POIs on those regions. The number of POIs in each region is codified in the colorbar, where a logarithmic scale was chosen to show the wide range of values. The great coverage of this dataset is clear, as well as the existence of regions with high concentrations of POIs reaching up to 70,128 POIs in a single square defined above. The plot is the result of projecting the coordinates of the globe on a rectangle which implies spatial distortions; still, the frontiers of continents are identifiable even though no borders were used or drawn. A high density of POIs can be appreciated in North America and Europe, but active regions are also present in South America, the Middle East, and East Asia. Although to a much lesser extent, there are some localized regions with high numbers of POIs in Africa and Oceania. The details of this information are discussed in the methods section in Table \ref{tab_countries}, where the 15 countries with most of the POIs are listed; 80\% of the venues belong to these countries.
\subsection*{Temporal activity of users in Foursquare}
\begin{figure}[t!]
    \centering
    \includegraphics[width=1.0\linewidth]{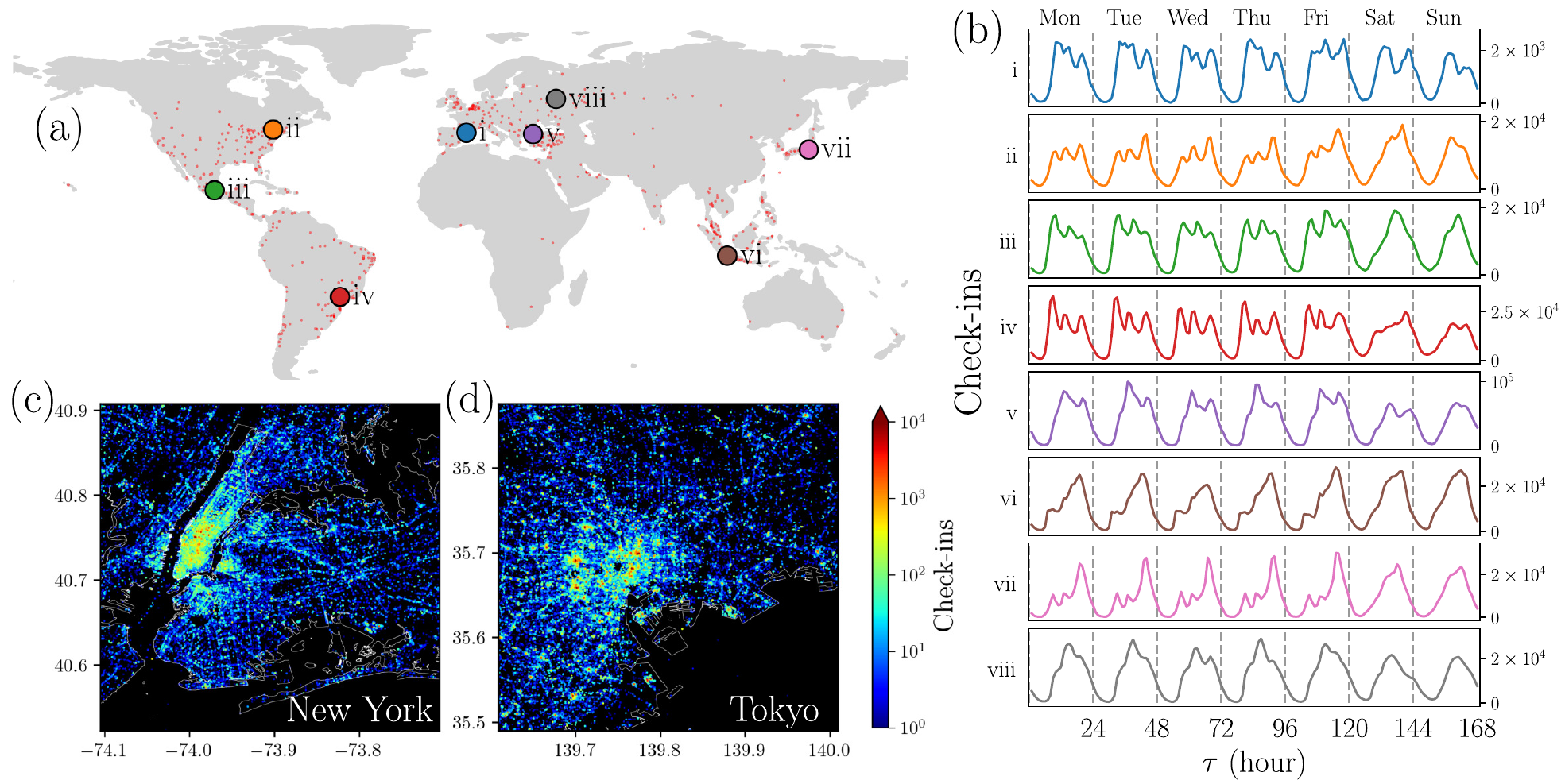}
    \caption{ {\bf Spatial and temporal analysis of  check-ins in different cities.} {(a) 632 cities worldwide analyzed, represented with small dots, and eight selected cities: (i) Barcelona, (ii) New York, (iii) Mexico City, (iv) Sao Paulo, (v) Istanbul (vi), Jakarta, (vii) Tokyo, and (viii) Moscow. (b)  Frequency counts for all the check-ins in the cities (i) to (viii) presented in (a) gathered using the local time when they were made. The counts reported in the histograms correspond to the number of check-ins in each hour of the week, from the first hour on Monday to the last hour on Sunday. (c) Points of interest in the Foursquare dataset in New York and (d) Tokyo. In this representation, the number of check-ins in the respective POI is codified in the colorbar. All figures were created using python 3.8 and the matplotlib (3.5.0) package (\url{https://matplotlib.org}). The map in panel (a) was created using the geopandas (0.12.1) package (\url{https://geopandas.org}). Land borders in panels (c) and (d)  were obtained from Open Street Maps Boundaries webpage (\url{https://osm-boundaries.com/})  }
    }
    \label{Fig_2}
\end{figure}
The classification of POIs and check-ins at the level of cities allows us to perform temporal, spatial and spatio-temporal analysis of features. For example, in Ref.\cite{riascos2017emergence} a detailed study of the activity in Foursquare, of users in New York and Tokyo, is presented. {We focused our study on cities that hold the majority of the data. From the total number of points of interest in the database, we selected only those belonging to cities with more than 10,000 check-ins. This resulted in a set of 632 cities located in 87 countries, as illustrated in Fig. \ref{Fig_2}. These 632 cities, represented by red dots on the map in panel \ref{Fig_2}(a), are located in countries across all continents, reflecting the diversity in social, economic, cultural, linguistic, and other terms, crucial for the study of human activity in cities. As we will see later, this diversity allows us to identify commonalities among all cities, but also and especially differences in activity patterns of cities. To illustrate these commonalities and contrasts, we selected eight cities that capture some of this diversity. These cities are Barcelona (Spain), New York (United States), Mexico City (Mexico), Sao Paulo (Brazil), Istanbul (Turkey), Jakarta (Indonesia), Tokyo (Japan), and Moscow (Russia). These cities are shown with different colors on the map in \ref{Fig_2}(a). In Fig. \ref{Fig_2}(b), we present the temporal distribution of check-ins made during the nearly 22 months of observation in these eight cities that were intentionally selected to represent the behaviors in different regions with a great variety of cultural and linguistic characteristics. Also,  these  cities were chosen because they have a greater number of check-ins.} Temporal information of the check-ins was grouped by urban area and was generated using the local time, obtained through the Universal Coordinated Time and adding the minutes corresponding to the timezone correction (see methods section for details). In this way, we can identify daily routines for many geographical areas around the world. For this study, we defined the time granularity as 168 hours (corresponding to a full week), so we have, for each region, a histogram with all the check-ins made in the period of observation gathered by hour. We considered this histogram as a characteristic print of human  temporal activity within a region. In doing so, clear patterns emerge with slight differences between regions. Some regularities are evident: there is low activity of check-ins during the night and high activity during the day; this matches the behavior reported by Yang, et al.\cite{yang2019revisiting}, for a subset of this data at a global scale. In this case, the pattern persists at a local scale, as has been shown for other phenomena\cite{prieto2021heartbeat,song2010limits, sparks2020global}. But, at this level, differences in patterns become noticeable; the part of the day that concentrates most of the activity, the maximum number of check-ins in a day, and the change of the activity patterns for weekdays and weekends describes the collective behavior of city inhabitants that varies from one city to another.
\\[2mm]
{In Figs. \ref{Fig_2}(c)-(d), we show the spatial distribution of POIs for two cities: New York and Tokyo. When placing a point at the coordinates of each POI, different spatial patterns emerge directly linked to urban infrastructure in the first place. Street grids and blocks in urban areas can be identified. But some patterns emerge, which are very different in each case, related to the quantity and density of the POIs. For example, a large area of high density can be seen in the case of New York, in the Manhattan area, while in Tokyo the high-density areas form more compact and dispersed clusters in a larger region. In addition, the attractiveness (total number of check-ins in a POI) of the venues is depicted in the colorbar that codifies the number of check-ins.}
The statistical analysis of the number of check-ins at each POIs reveal characteristics of a complex urban environment with a power-law behavior in which most POIs register a lower number of check-ins while a few POIs have an attractiveness several orders of magnitude greater. Again, a similar result has been reported by Yang et al.\cite{yang2019revisiting}, at a global scale for the same dataset. This power-law behavior in the attractiveness of POIs is also observed at different scales in other studies; for example, in the attractiveness of sites measured from the activity of taxis in New York City \cite{riascos2020networks} or the importance of airports in the United States \cite{Ruiz_2022_airports}. In this context, our findings in  Fig. \ref{Fig_2}(c)-(d) show that the activity of users in Foursquare is associated with the complexity in the distribution of POIs. This feature is observed in all the cities in this study.
\\[2mm]
\begin{figure}[t!]
\centering
\includegraphics[width=0.85\linewidth]{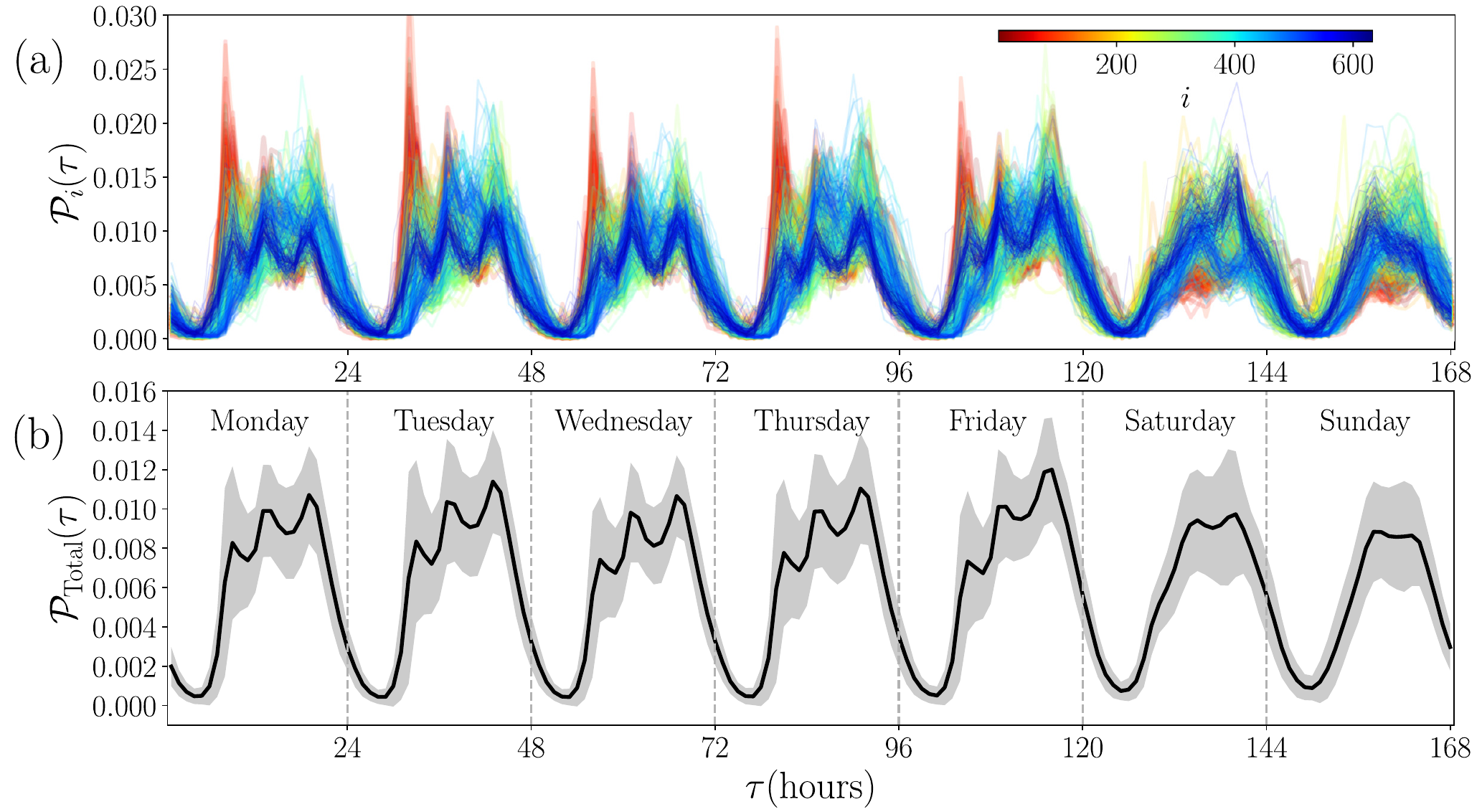}
\caption{{\bf Temporal analysis of check-ins in 632 cities.} (a) Probability  $\mathcal{P}_i(\tau)$ of check-in at times $\tau$ in  $N=632$ cities denoted by $i=1,2,\ldots, N$. All the check-ins in each city were gathered by the local time when each register was made. Frequency counts are normalized over the week to obtain each $\mathcal{P}_i(\tau)$. The time $\tau$ can take the values from 1 (the first hour of Monday) to 168 (the last hour of Sunday), and the results for each city $i$ (codified in the colorbar) are presented with continuous lines. (b) The probability  $\mathcal{P}_{\mathrm{Total}}(\tau)$ obtained for all the check-ins in the 632 cities is plotted with a continuous line. In this case, the standard deviation $\sigma(\tau)$ for all the check-ins and all the cities in (a) evaluated at time $\tau$ defines variations of this probability. The results are shown as a region defined by $\mathcal{P}_{\mathrm{Total}}(\tau)\pm \sigma(\tau)$ presented with gray color. All figures were created using python 3.8 and the matplotlib (3.5.0) package (\url{https://matplotlib.org}). }
\label{Fig_3}
\end{figure}
{In Fig. \ref{Fig_3} we extend the temporal analysis of check-ins to $N=632$ cities in $87$ countries. We consider activity counts as in panels in Fig. \ref{Fig_2}(b),} all the information of check-ins in a city $i$ were gathered using the corresponding local time $\tau$. The activity counts are normalized over the week to obtain the relative frequency or probability $\mathcal{P}_i(\tau)$. The time $\tau$ can take the values from 1 (associated to the first hour of Monday) to 168 (the last hour of Sunday). In Fig. \ref{Fig_3}(a), we present the probability  $\mathcal{P}_i(\tau)$ of check-in at times $\tau$ in  cities denoted by $i=1,2,\ldots, 632$. The results reveal marked differences between low nocturnal activity and high daytime activity. Also, patterns emerge in all of them by grouping check-ins by their local time $\tau$ of occurrence. For all the cities, fluctuations in the values of   $\mathcal{P}_i(\tau)$ are small at night, but during the day we see different behaviors of the curves describing each city with all kinds of deviations; for example, in some cities  $\mathcal{P}_i(\tau)$ changes considerably in the early hours of the morning, others at night, whereas other cities differ considerably at weekends. To show these variations more clearly, in Fig \ref{Fig_3}(b) we depict  $\mathcal{P}_{\mathrm{Total}}(\tau)$, obtained for all the check-ins in the $N=632$ cities. This quantity is equivalent to calculate the average between all cities
\begin{equation}
\mathcal{P}_{\mathrm{Total}}(\tau)=\frac{1}{N}\sum_{i=1}^N \mathcal{P}_i(\tau).
\end{equation}
On the other hand,   the respective standard deviation $\sigma(\tau)$ of the values $\mathcal{P}_i(\tau)$ give us a measure of the differences observed in the cities considered. Our findings are shown in Fig. \ref{Fig_3}(b) as a shaded region defined by $\mathcal{P}_{\mathrm{Total}}(\tau)\pm \sigma(\tau)$. The dispersion of the values in particular hours of the day can be seen, and noticeable variations between weekdays and weekends can be observed.
\begin{figure}[b!]
\centering
\includegraphics[width=.8\linewidth]{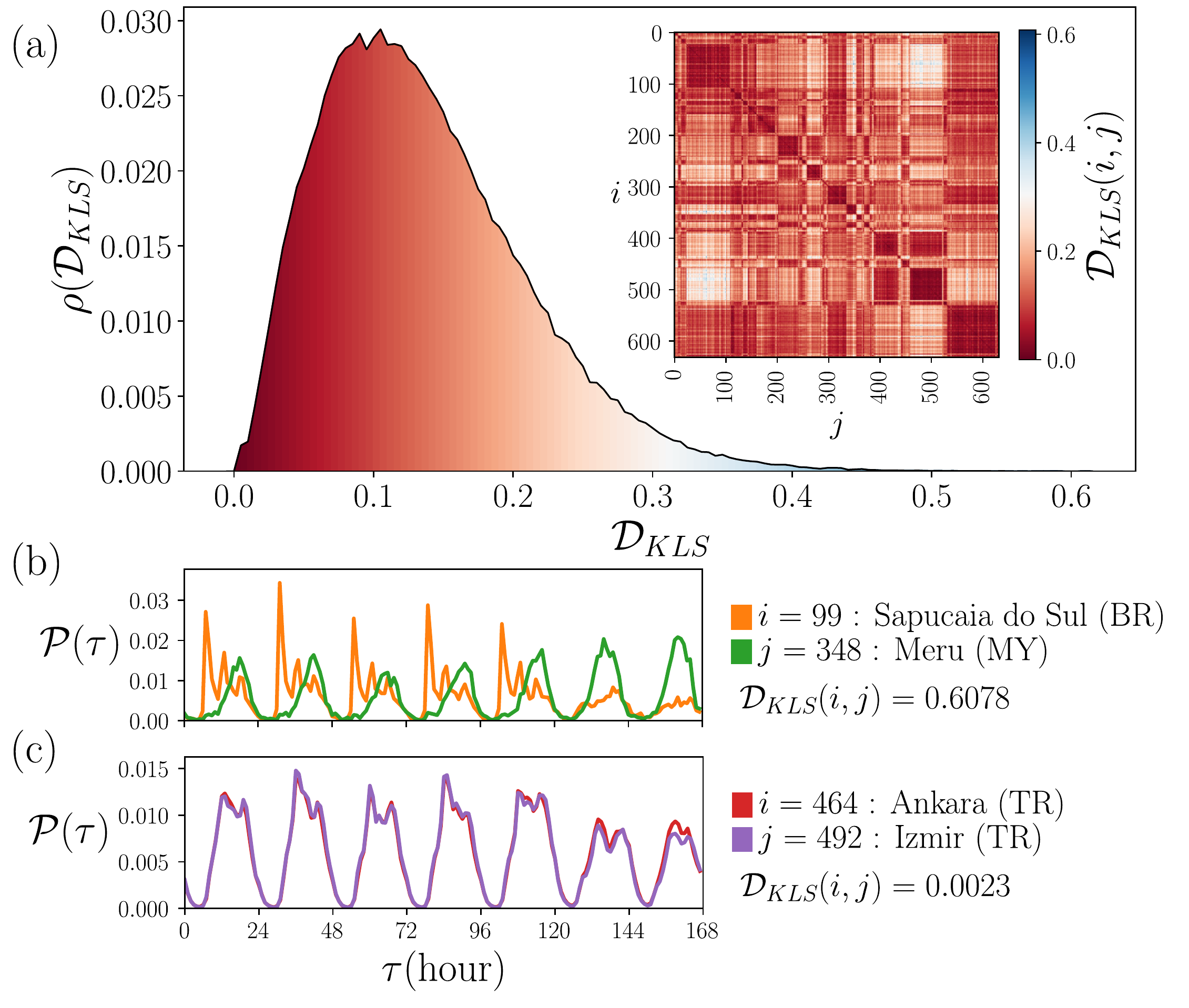}
\vspace{-2mm}
\caption{{\bf Comparison of temporal activity between cities.} (a) Statistical analysis of the symmetric Kullback-Leibler distances $\mathcal{D}_{KLS}$ from the comparison of check-ins temporal distribution by pairs. The values are obtained from Eq. (\ref{eq:KLS}) for all the city pairs $i,~ j~ =~ 1,\ldots,~ 632$. The probability density $\rho (\mathcal{D}_{KLS})$ is obtained using bin counts in intervals with $\Delta \mathcal{D}_{KLS} = 0.005$.  The matrix with all the elements $\mathcal{D}_{KLS}(i,j)$ is presented as an inset. (b) The two most different probability densities of the set are of the cities Sapucaia do Sul in Brazil and Meru in Malaysia, with $\mathcal{D}_{KLS}=0.6078$; the maximum value found. (c) The two most similar cities of our sample, are Ankara and Izmir both in Turkey, with $\mathcal{D}_{KLS}=0.0023$, the minimum non-null value. All figures were created using python 3.8 and the matplotlib (3.5.0) package (\url{https://matplotlib.org}). }
\label{Fig_4}
\end{figure}
\subsection*{Comparison of temporal activity between cities using the Kullback-Leibler divergence}
{
The variety of results observed for the probabilities  $\mathcal{P}_i(\tau)$ in Fig. \ref{Fig_3}(a) motivates the exploration of a criterion to compare the temporal activity between two particular cities $i,j$. To this end, we use the Kullback-Leibler divergence, also known as relative entropy, defined by \cite{KullbackLiebler_1951, CoverThomasBook}
\begin{equation}\label{eq:KL}
\mathcal{D}_{KL}(i,j) = \sum_{\tau}\mathcal{P}_i(\tau)\log \left[ {\frac{\mathcal{P}_i(\tau)}{\mathcal{P}_j(\tau)}} \right],
\end{equation}
where the sum in time $\tau$ ranges from $1$ to $168$ (all the hours in a week) and $i,j = 1,2,\ldots,N$ (see the methods section for details). The Kullback-Leibler divergence satisfies our interest to compare the temporal distributions of check-ins, but its definition does not generate a symmetric measure since $\mathcal{D}_{KL}(i,j)$ is different from $\mathcal{D}_{KL}(j,i)$. This would cause that even if a city $i$ is determined to be similar to another city $j$, $j$ is not necessarily similar to $i$. However, the average of the Kullback-Leibler divergence between pairs $(i,j)$ and $(j,i)$
\begin{equation}\label{eq:KLS}
  	  \mathcal{D}_{KLS}(i,j)\equiv \frac{\mathcal{D}_{KL}(i,j)+\mathcal{D}_{KL}(j,i)}{2}
\end{equation}
is symmetric. In this manner, similar distributions produce a small value and dissimilar ones are associated with larger values. Then, this symmetric quantity is adequate to describe the similarity between the temporal distributions of cities.
}
\\[2mm]
In Fig. \ref{Fig_4}, we present the results obtained from the evaluation of $\mathcal{D}_{KLS}(i,j)$, in Eq. (\ref{eq:KLS}), to compare the temporal activity presented in Fig. \ref{Fig_3}(a) for all the  cities $i,j=1,2,\ldots, 632$, considering the information of user's check-ins in Foursquare. In Fig. \ref{Fig_4}(a) we present the statistical analysis of  $\mathcal{D}_{KLS}(i,j)$ for all pairs of cities. The results are depicted as a probability density $\rho(\mathcal{D}_{KLS})$ for the values $\mathcal{D}_{KLS}$, that is, we calculate the values of $\mathcal{D}_{KLS}(i,j)$ for all the pairs $(i,j)$. With these values, we obtained the histogram shown. We show, as an inset, the $N\times N$ matrix with elements $\mathcal{D}_{KLS}(i,j)$; the respective values are codified in the colorbar. The results for $\rho(\mathcal{D}_{KLS})$ show that a high fraction of the entries in the matrix have values $\mathcal{D}_{KLS}$ around $0.1$. For the diagonal elements we have $\mathcal{D}_{KLS}(i,i)=0$. On the other hand, the maximum value for two different cities is $\mathcal{D}_{KLS}=0.6078$, and occurs between the Brazilian city of Sapucaia do Sul and Meru, in Malaysia, as shown in  Fig. \ref{Fig_4}(b). At the opposite extreme, the minimum non-null value reached is $\mathcal{D}_{KLS}=0.0023$, found for the comparison between Ankara and Izmir, both in Turkey. The respective probabilities are shown in  Fig. \ref{Fig_4}(c). The results observed in Fig. \ref{Fig_4}(b) are reasonable since we are comparing the activity in two complete different urban areas.  In contrast, for the cities considered in Fig. \ref{Fig_4}(c), the resemblance between these two cities is remarkable considering that they are more than 520 km apart. Additionally, a review of the data shows that of the 82,285 active users in Ankara and the 90,923 active users in Izmir, only 11,141 made check-ins in both cities; this represents only 6.87\% of the total users with activity in these regions. The similarity in their patterns is not explained by common users but by comparable urban behavior in the same country.

\subsection*{Networks and temporal patterns between cities}

In this section, we apply methods of network science to analyze the similarities between cities. To this end, we define a network in which nodes represent cities and links the similarity relationship between cities. In this way, two nodes are connected if the respective cities have similar temporal activity. To generate this structure, it is necessary to define what is considered sufficiently similar. We use a threshold value $H$. If two cities have values  $\mathcal{D}_{KLS}$ in Eq. (\ref{eq:KLS}) lower or equal than $H$ then these cities are considered similar. All this information defines a similarity network for each value $H$. The respective $N\times N$  adjacency matrix is denoted as $\mathbf{A}(H)$, with elements $i$, $j$ given by
\begin{equation}
    A_{ij}(H) = \left\{
    \begin{array}{ll}
         1 \qquad & \mathcal{D}_{KLS} (i, j) \leq H,\\
         0 & \mathrm{otherwise}.
    \end{array}
    \right.
    \label{adjacency}
\end{equation}
Additionally, we require $A_{ii}(H)=0$ for $i=1,2,\ldots,N$, to avoid self loops. From the symmetry of the distance $\mathcal{D}_{KLS}$, follows the symmetry of $\mathbf{A}(H)$, defining an undirected network.
\\[2mm]
\begin{figure}[!t]
\centering
\includegraphics[width=.95\linewidth]{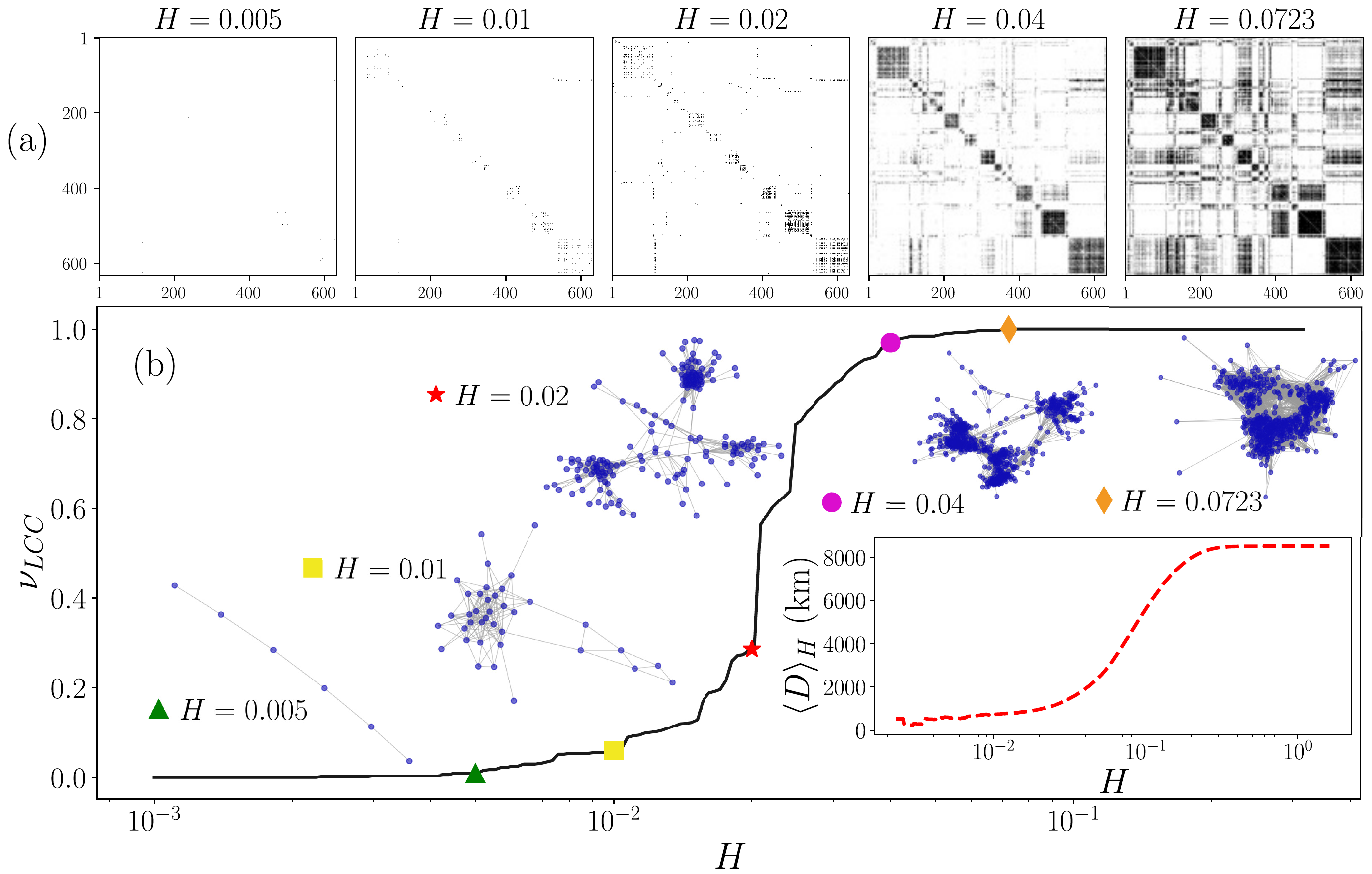}
\vspace{-2mm}
\caption{{\bf Similarity networks generated using different threshold values $H$.} (a) Adjacency matrices  $\mathbf{A}(H)$ with elements given by Eq. (\ref{adjacency}) using the  values $H\in \{0.005,\, 0.01,\, 0.02,\, 0.04,\, 0.0723\}$, binary entries $A_{ij}(H)$ are depicted in white for 0 and black for 1. (b) Fraction of nodes in the largest connected component $\nu_{LCC}$ as a function of $H$ in the interval $0.001\leq H \leq  0.32$. The largest connected component of the networks generated with $\mathbf{A}(H)$ in panel (a) are presented and the inset shows $\langle D \rangle_H$ calculated using Eq. (\ref{Distance_D_H_geo}). All figures were created using python 3.8 and the matplotlib (3.5.0) package (\url{https://matplotlib.org}). 
}
\label{Fig_5}
\end{figure}
In Fig. \ref{Fig_5}, we analyze similarity  networks as a function of $H$. In Fig. \ref{Fig_5}(a) we depict the adjacency matrix $\mathbf{A}(H)$ for different values of $H$. In this manner, for each $H$ all the information of $\mathcal{D}_{KLS}(i,j)$ in Fig. \ref{Fig_4}(a) is converted into a binary matrix with entries 0 and 1. In Fig.  \ref{Fig_5}(b), we present the fraction of nodes that belong to the Largest Connected Component, $\nu_{LCC}=S_{LCC}/N$, where $S_{LCC}$ is the size of the  Largest Connected Component (LCC). The LCC obtained for the values of $H$ explored in Fig. \ref{Fig_5}(a) are shown as insets.
\\[2mm]
In the results in Fig. \ref{Fig_5}(b), it is worth noticing that for $H= 0.005$, the network is formed by disconnected subnetworks with a few nodes and the LCC is a linear graph with 6 nodes; for $H< 0.0023$ each node is disconnected. On the other hand, the network is fully connected for $H>0.6078$; the maximum value of $\mathcal{D}_{KLS}$ found. The transition between these two limits gives rise to a convenient choice of $H$ corresponding with a network that captures the nature of the similarity between the cities that we are analyzing. The results for $\nu_{LCC}$ reveal that the size of the LCC contains more than 90\% of the nodes at $H=0.031$, 99\% at $H= 0.052$. On the other hand, $H=0.0723$ is the minimal value of the similarity threshold that produces a connected network that includes all the $N=632$ nodes (cities). Finally, we see that around $H=0.02$, $\nu_{LCC}$ suffers an abrupt change with $H$ that is analogous to a percolation threshold \cite{BarabasiBookNS}, separating two regimes: one with   $\nu_{LCC}<0.3$, defining small sub-networks with high similarity in the temporal information and a second one with $\nu_{LCC}>0.6$ where the connected networks incorporate a high fraction of the cities. 
\\[2mm]
{
On the other hand, in order to explore the relationship between the edges in each network generated with the value $H$ and the geographical distance between cities, we define
\begin{equation}\label{Distance_D_H_geo}
	\langle D\rangle_H\equiv\frac{\sum_{i,j=1}^N A_{ij}(H)d_{ij}}{\sum_{i,j=1}^NA_{ij}(H)}=\frac{\sum_{i,j=1}^N A_{ij}(H)d_{ij}}{2\,\mathcal{E}(H)},
\end{equation}
where $d_{ij}$ denotes the geographical distance between cities $i$ and $j$ and $\mathcal{E}(H)$ is the number of edges of the graph associated with $\mathbf{A}(H)$ (see methods section for details on the calculation of geographical distances between cities). In Fig. 5(b), we present as an inset the value $\langle D\rangle_H$ as a function of $H$. The results show that for small $H$, cities having very similar temporal activity histograms are also geographically closer. These average distances remain relatively low up to the network with $H=0.05$. However, after this value, the average distances grow to $8512$ km which is the average distance between all the cities analyzed.}
\\[2mm]
\begin{figure}[!t]
\centering
\includegraphics[width=1.0\linewidth]{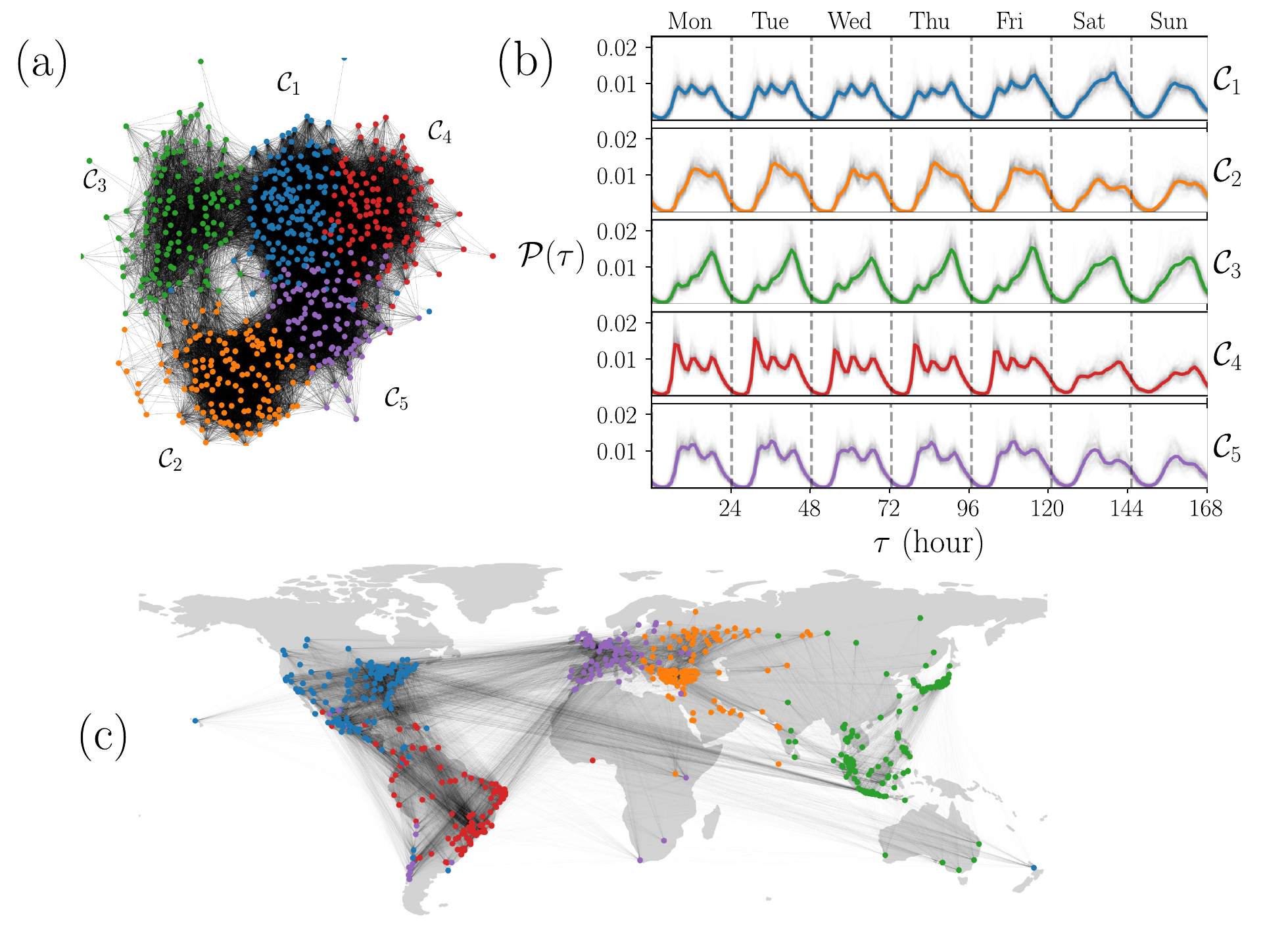}
\vspace{-8mm}
\caption{{\bf Human activity patterns identified using community detection in similarity networks}. (a) Community structure of the similarity network generated with $H=0.0723$, five communities $\mathcal{C}_1$, $\mathcal{C}_2,\mathcal{C}_3,\mathcal{C}_4$, $\mathcal{C}_5$ were detected using the Louvain's algorithm and are represented with different colors. (b) Probability $\mathcal{P}(\tau)$ for the temporal activity of the cities in each community. Thin gray lines present the curves $\mathcal{P}_i(\tau)$ depicted in Fig. \ref{Fig_3}(a) whereas the colored thick line represents the statistical analysis of all the check-ins in the cities of the community. (c) Geographical representation of the network in (a). In this case, each node is a city depicted on a world map. All figures were created using python 3.8 and the matplotlib (3.5.0) package (\url{https://matplotlib.org}). The map in panel (c) was created using the geopandas (0.12.1) package (\url{https://geopandas.org}).}
\label{Fig_6}
\end{figure}
Once we established a criteria to build similarity networks using the temporal activity of users of Foursquare in Fig. \ref{Fig_3}(a), we can apply community detection algorithms. The concept of community has emerged in network science as a method for finding groups within complex systems identifying sub-networks with statistically significantly more links between nodes in the same group than nodes in different groups \cite{GirvanNewmanPRE2004,NewmanPNAS2006,Fortunato2010Community}. In our similarity network, these communities represent groups of cities with comparable activity $\mathcal{P}_i(\tau)$. In Fig. \ref{Fig_6} we present the results for a network with $N=632$ nodes generated with $H=0.0723$.
\\[2mm]
In panel \ref{Fig_6}(a) we depict five communities $\mathcal{C}_1$, $\mathcal{C}_2, \ldots,\mathcal{C}_5$ detected using the Louvain's algorithm \cite{Blondel_2008} implemented in the library NetworkX in Python \cite{NetworkX}, {see methods section for a description of this algorithm.} The number of nodes on each community $\mathcal{C}_s$ (with $s=1,2\ldots,5$) is denoted by $\mathcal{N}_s$ and, from this analysis we obtain groups of cities with $\mathcal{N}_1=148$, $\mathcal{N}_2=136$, $\mathcal{N}_3=133$, $\mathcal{N}_4=113$, and $\mathcal{N}_5=102$. In Fig. \ref{Fig_6}(b) we plot with thin gray lines the probabilities $\mathcal{P}_i(\tau)$ showed in Fig. \ref{Fig_3}(a) in groups defined by each community $\mathcal{C}_s$; each panel contains $\mathcal{N}_s$ curves. In addition, we include the statistical analysis considering all the check-ins in each group; the results are shown with colored thick lines.  When grouped in this way, the curves observed within each community $\mathcal{C}_s$ are similar,  evidencing the fact that they have the same shape as the averages. 
\\[2mm]
The average curves in Fig. \ref{Fig_6}(b) show that the $\mathcal{C}_1$ community, whose average is plotted in blue, has a behavior from Monday to Friday characterized by three peaks throughout the day (at 8, 12 and 18 hours). On weekends, this pattern is broken and a single maximum is observed around 20 hours on Saturday and at noon on Sunday. The $\mathcal{C}_2$ community, with the orange average line, is characterized by a pronounced maximum at 13 hours and a second relative maximum at 19 hours, from Monday to Friday. This behavior is maintained on weekends but with less check-ins. The community $\mathcal{C}_3$, with a curve  in green, is the one with the least contrast between the shape from weekdays versus weekends. Every day the maximum is found at 18 hours. Still, from Monday through Friday there is a small local maximum at 8 hours that disappears on the weekend. $\mathcal{C}_4$ with the average presented in red, has the same features as $\mathcal{C}_1$ but with different relative sizes; in this case, the first daily maximum dominates over the rest. The $\mathcal{C}_5$ community is the smallest and whose average behavior is depicted in purple. In this case the curve suffers different changes throughout the week. While there is a pattern with maxima at 9, 12 and 19 hours, and a valley at 15 hours, from Monday to Friday, the relative sizes are not always equal; on Monday the first and second peaks dominate, on Tuesday the second maximum is the largest, on Wednesday the three are practically the same size, on Thursday the first two peaks almost merge and dominate over the third, and on Friday the first almost disappears while the second peak dominates. Finally, on Saturday there is a change giving rise to two maximums at 13 and 20 hours, also present on Sunday although smaller.
\\[2mm]
In addition, considering that each probability curve $\mathcal{P}_i(\tau)$ corresponds to a node $i$, which in turn is a city with given coordinates, in Fig. \ref{Fig_6}(c) we plot the network with each node located in a world map. The node colors and the whole network connectivity are the same as in Fig. \ref{Fig_6}(a). This network representation shows that the communities formed from the similarity of city behaviors correspond to well-defined geographic regions. Cities in North America belong predominantly to the $\mathcal{C}_1$ community; those in Eastern Europe and the Middle East belong to  $\mathcal{C}_2$; the $\mathcal{C}_3$ community is composed of several cities in Eastern India, and cities in East Asia and Oceania; $\mathcal{C}_4$ contains most cities in South America, and most of the cities in Western Europe belong to the $\mathcal{C}_5$ community. This is a remarkable result that we will discuss further below.
\\[2mm]
Once we defined the communities of the network, we can compare the fraction of intra-community and inter-community links. Defining $L_s$ as the number of links between nodes in community $\mathcal{C}_s$, and $L_{st}$ as the number of links between a node in $\mathcal{C}_s$ and a node in $\mathcal{C}_t$, we calculate the fraction of inter-community links as $L_{st} /(\mathcal{N}_s\mathcal{N}_t)$ and intra-community links as $2 L_s/ (\mathcal{N}_s(\mathcal{N}_s -1))$. The values obtained allow us to compare the number of links with the total number of possible links. The values for the fraction of intra-community links are $0.79$, $0.68$, $0.51$, $0.76$ and $0.75$ for $\mathcal{C}_1,\ldots,\mathcal{C}_5$, respectively. Regarding the fraction of inter-community links, the values are below $0.1$ except between communities $\mathcal{C}_1$ and $\mathcal{C}_4$ (North America and South America) with $0.32$ and, with $0.26$, between $\mathcal{C}_1$ and $\mathcal{C}_5$ (North America and Europe); there are a fraction $0.15$ of inter-community nodes between $\mathcal{C}_2$ (Middle East) and $\mathcal{C}_5$ (Europe), and $0.12$ between $\mathcal{C}_5$ and $\mathcal{C}_4$ (South America).  The results also show that $\mathcal{C}_3$ has little similarity with the other communities. Europe is a community with many elements in common with America and the Near East. East Asia, on the other hand, has few elements in common with the rest of the world in terms of temporal patterns of human behavior. These features are observed in the map in Fig. \ref{Fig_6}(c) evidencing cultural and historical features of each region. 
\\[2mm]
Finally, it is worth mentioning that, unlike other cluster identification methods, e.g., agglomerative clustering (discussed in the next section), information on similarity between specific regions within the same communities is preserved in the individual links. In Fig. \ref{Fig_6}(c), it is clear the high density of links between North America and the United Kingdom, as well as Brazil and Portugal, regions with cultural and linguistic  relationships. The results also show that not all cities belong to the same community as would be expected based on their geographic region. Specifically, there are 8 cities in Chile, 3 in Mexico and 1 in Uruguay that are more similar to European cities than to American ones.
\subsection*{Identification of patterns using machine learning}
\begin{figure}[t]
\centering
\includegraphics[width=1.0\linewidth]{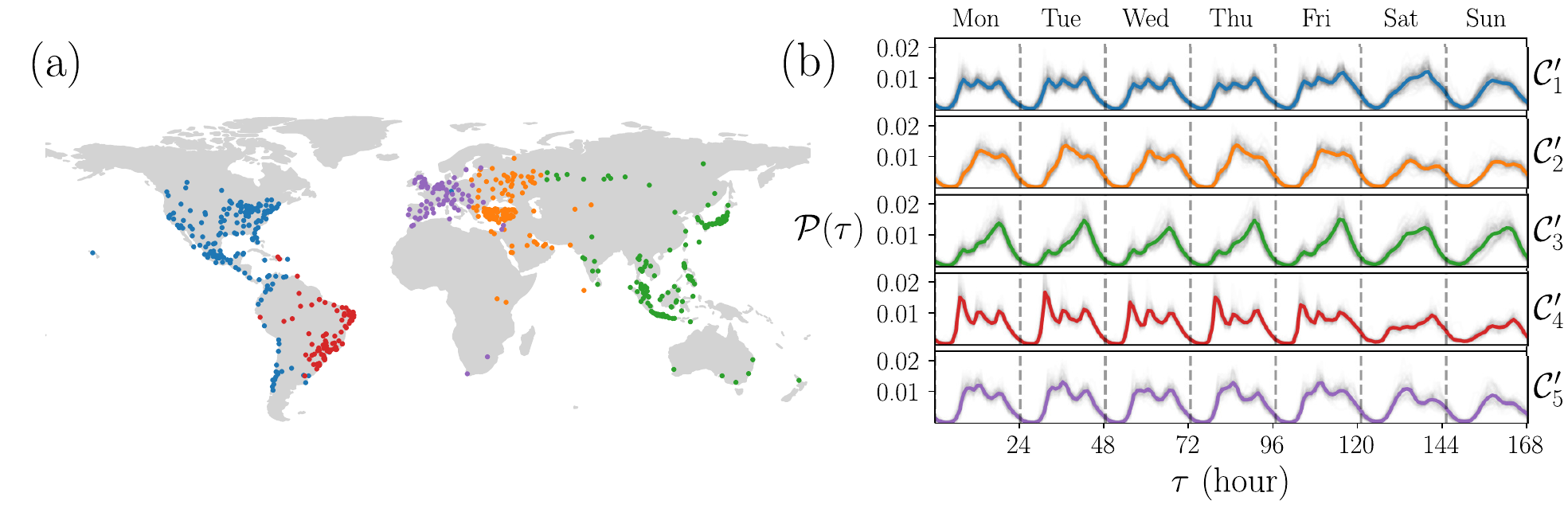}
\caption{\label{Fig_7} Pattern identification using agglomerative hierarchical clustering. (a) Geographical representation of 5 clusters detected. (b) Activity patterns in each group. All figures were created using python 3.8 and the matplotlib (3.5.0) package (\url{https://matplotlib.org}). The map in panel (a) was created using the geopandas (0.12.1) package (\url{https://geopandas.org}).}
\end{figure}
Machine learning can be used as an alternative approach, with many algorithms involved, to classify objects or patterns in a very efficient way \cite{MLBook1,Agglomerative1}. In particular, we used the unsupervised hierarchical agglomerative clustering to classify our temporal patterns of activity in cities. When we apply this algorithm to our data set, we obtained five clusters, as shown in Fig. \ref{Fig_7}. This number of clusters or communities is obtained by maximizing the modularity. The clusters of cities detected are presented in Fig. \ref{Fig_7}(a). For comparison, these clusters are depicted using the same colors as those used in Fig. \ref{Fig_6}(c). In this classification of the dataset of check-ins, each cluster obtained by this procedure is denoted as $\mathcal{C}^\prime_s$ and contains $\mathcal{N}^\prime_s$ elements (with $s=1,2\ldots,5$). We have: $\mathcal{N}^\prime_1 = 178$, $\mathcal{N}^\prime_2 = 128$, $\mathcal{N}^\prime_3 = 140$, $\mathcal{N}^\prime_4 = 90$, and $ \mathcal{N}^\prime_5 = 87$. A comparison of Fig. \ref{Fig_7}(a) and our findings in Fig. \ref{Fig_6}(c) shows that the differences between $\mathcal{N}_s$ and $\mathcal{N}_s'$ are mainly due to cities from South America and the Caribbean that were in $\mathcal{C} _1$ (blue) are now in $\mathcal{C}^\prime_4$ (red), and between Russian and Indian cities that were in $\mathcal{C}_2$ (orange) now belong to $\mathcal{C}^\prime_3 $ (green). In addition, Fig. \ref{Fig_7}(b) shows the probabilities $\mathcal{P}_i(\tau)$ grouped according to this classification. Again, in each panel, corresponding $\mathcal{C}^\prime_s$, the $\mathcal{N}^\prime_s$ curves of the cities are shown in thin gray, while the average behavior for each time $\tau$ is shown with a colored thick line. The resemblance to the behaviors shown in Fig. \ref{Fig_7}(b) is remarkable showing that the emergent patterns in the dataset can be detected by different algorithms. 
\section*{Discussion}
In this work, we use Foursquare check-ins as a proxy of human activity in cities. The data set explored provides vast information about people's interests, site characteristics, and behavioral patterns in cities, among many others.  The information analyzed contains 90,048,627 check-ins made by 2,733,324 users from April 2012 to January 2014 with active check-ins obtained from registers on Twitter by searching a hashtag generated automatically for users that linked their Foursquare activity with that social network. Information from 632 cities from 87 countries, with more than 10,000 POIs, was used. We explore the spatial properties of POIs and check-ins and the temporal features of the check-ins which, as self-reported spatio-temporal interactions between people and places, provide valuable information about activities carried out in cities. In particular, we analyze statistically the check-ins per hour on  weeks for each city.
\\[2mm]
From the probabilities describing the temporal activity of each city, we apply a symmetrized  Kullback-Leibler divergence to compare these probabilities across all 632 cities. From this information, we define similarity networks in terms of a threshold value $H$ to decide if two cities have a similar activity or not. We explore the size of the largest connected component as a function of $H$; in particular, we found that around $H=0.02$ a critical percolation threshold exists, whereas for $H=0.0723$ the largest connected component includes all the nodes of the network. Our findings reveal collective emergent behaviors that go beyond the spatial and mobility aspects that define metropolitan areas, megacities, or functional urban areas, and allows for the tracking of higher-order structures, as was proven in some cities that have great similarity, despite the fact that their geographical separation prevents them from being the same.
 \\[2mm]
We apply Louvain's algorithm for community detection to the similarity network generated with $H= 0.0723$; the results define 5 groups of cities with similar activity of check-ins. By locating all the cities in a map, we notice a remarkable pattern: The five communities of cities belong to five different regions that correspond to different continents. In addition, we use a Machine Learning algorithm to classify the activity patterns obtained. Although this method is completely different from the Network Science approach, we nevertheless obtained basically the same classification. Specifically, the Machine Learning algorithm that we used was the Unsupervised Agglomerating Clustering \cite{MLBook1,Agglomerative1}. Both approaches, although very different, lead to the same classification of five different communities worldwide. Even more surprising, these five communities clearly correspond to five different regions on a planetary scale: 1) North America, (2) South America, (3) West Europe, (4) East Europe and Middle East, and (5) East Asia.   
\\[2mm]
{
It can be clearly noticed that practically in all cities there are these patterns of maxima and minima of human activity at the same hour during the week. At the same time, we notice a different pattern during the weekdays and another one slightly different during weekends. These patterns are very robust in the sense that the maxima and minima are clearly noticeable, even in the average curves involving the 632 cities in the analysis. 
Since these 632 cities are distributed in many parts of the world, spanning 5 continents, we think that the patterns are related to some universal aspects of human behavior. For instance, the fact that we humans are universally a diurnal species. This is one of the so-called human universals in the literature of human evolution. This fact can be seen in the minima of activity for all cities at deep night (around of 2 or 3 am). This universal periodicity is also related to circadian rhythms in humans and other internal clocks studied in chronobiology. However, the 3 peaks during the day, which appear early in the morning (between 6 and 7 am), at noon (around 12 or 13 hours), and in the afternoon (around 19 and 20 hours), can be related more with modern routines associated with working hours.
}
\\[2mm]
In summary, using a vast data set from location-based social networks, we analyze 632 cities of 87 countries around the world to obtain temporal patterns of human activity that characterize points of interest within the cities. Using network science and machine learning algorithms we unravel communities with different patterns of human  behavior. This suggests that human activity patterns can be universal with some geographical, historical and cultural variations on a planetary scale.    

\section*{Methods}
{
\subsection*{Foursquare data and demography}
In this section, we recall some recent demographics of Foursquare. The user demographics is the following\cite{financesonline}:
\begin{enumerate}
\item[] $\bullet$ urban communities: 64\%, towns and rural communities: 36\%.
\item[] $\bullet$ medium-sized town: 28\%, large city: 26\%
\end{enumerate}
This means that we have users of Foursquare from both urban, town and rural communities. Even though the urban percentage is larger, as expected, the difference is not that high; the rural or town percentage is roughly 1 in 3. This means that the data set used in our study reflects not only an urban feature but reflects small town or rural features as well.
\\[2mm]
On the other hand, the gender and age distribution is as follows\cite{similarweb}:
\begin{enumerate}
\item[] $\bullet$ male: 52\%, female: 48\%.
\item[] $\bullet$ ages: 18-24 years old: 19\%, 25-34 years old: 32\%,  35-44 years old: 20\%, 45-54 years old: 14\%, 55-64 years old: 10\%, 65+ years old:  6\%.
\end{enumerate}
Here we notice that the percentage of males and females is basically the same: half and half. Therefore, our data set and analysis can be applied to both females and males. Regarding age distribution, we notice that about half (50\%) of users of Foursquare have ages between 25 and 44 years old. This is expected since people at that age: young adults, are the ones that show more mobility and activity. It is clear that the mobility, activity and interest in reporting through check-ins to the social network tends to diminish with age. However, the distribution is somewhat broad.
\\[2mm]
Regarding the type of places visited and checked as POIs, it is worth mentioning that the categories are very broad. There are in Foursquare more than 1100 venue categories, distributed in ten major Foursquare categorization groups of points of interest POIs\cite{4S_categories}:
Arts and Entertainment, Business and Professional Services, Community and Government, Dining and Drinking, Event, Health and Medicine, Landmarks and Outdoors, Retail, Sports and Recreation, Travel and Transportation.
This points to a large diversity of interests that are captured in the data set that we used in our analysis.
\\[2mm]
Other relevant data about Foursquare are the following\cite{4S_countries, 4S_places}:
\begin{enumerate}
\item[] $\bullet$ Foursquare Places has over 100 million POIs across 247 countries and territories, as of January 4, 2023.
\item[] $\bullet$ 100M+ global POI
\item[] $\bullet$ 200+ countries and territories
\item[] $\bullet$ 14 billion user verified check-ins
\item[] $\bullet$ 1100+ venue categories
\item[] $\bullet$ 1 billion +  photos, tips, reviews
\end{enumerate}
Regarding the numbers of the particular data set used in this research, let us point out that we used data of Foursquare from April 2012 to January 2014. All the details of the data set and the processing, data mining, and data analytics, can be found in the following sections: Discussion, Methods – Dataset description, and Methods – Grouping POIs by urban area. Many of the results can be found in Tables \ref{tab_countries} and \ref{tab_cities}. Just to summarize the size of the data set studied, let us list the numbers involved:
\begin{enumerate}
\item[] $\bullet$ User: 2,733,324
\item[] $\bullet$ POIs: 11,180,160
\item[] $\bullet$ Check-ins: 90,048,627
\item[] $\bullet$ Countries: 253 (87 of which include 99\% of the data)
\item[] $\bullet$ Cities: 6,463 (632 with more than 10,000 check-ins)
\item[] $\bullet$ Continents: 5
\item[] $\bullet$ Categories of POIs: 519
\end{enumerate}
Given all these demographics together with the numbers, diversity, and vastness of the data set analyzed, and the use of aggregate data, the results can be a good description of the temporal activity in cities worldwide.
\\[2mm]
}

\subsection*{Dataset description}
\begin{table}[t]
\centering
\begin{tabular}{|c|l|c|c|c|c|c|}
\hline
\multicolumn{2}{|c|}{ Country} & \multicolumn{2}{c|}{POIs} & \multicolumn{2}{c|}{Check-ins} &  Users \\
\hline
Code & Name & Number & \% & Number & \% & Number \\
\hline
 US &   United States &  1990327 &  17.80 &  12778097 &  14.19 &  426341 \\
 ID &       Indonesia &  1198611 &  10.72 &   7765315 &   8.62 &  361193 \\
 BR &          Brazil &  1159258 &  10.37 &   9991354 &  11.10 &  261079 \\
 TR &          Turkey &  1098373 &   9.82 &  17500113 &  19.43 &  592630 \\
 RU &          Russia &   546532 &   4.89 &   4291601 &   4.77 &  122268 \\
 JP &           Japan &   519409 &   4.65 &   4784080 &   5.31 &   81293 \\
 MY &        Malaysia &   493453 &   4.41 &   4926145 &   5.47 &  127390 \\
 MX &          Mexico &   408434 &   3.65 &   3981409 &   4.42 &  147563 \\
 TH &        Thailand &   353444 &   3.16 &   2633608 &   2.92 &   82765 \\
 PH &     Philippines &   219097 &   1.96 &   1998063 &   2.22 &   60197 \\
 ES &           Spain &   212161 &   1.90 &   1083153 &   1.20 &   67638 \\
 GB &  United Kingdom &   210777 &   1.89 &   1271622 &   1.41 &   77949 \\
 IT &           Italy &   197007 &   1.76 &    867931 &   0.96 &   52394 \\
 CL &           Chile &   195226 &   1.75 &   2209981 &   2.45 &   53714 \\
 DE &         Germany &   142347 &   1.27 &    623759 &   0.69 &   45574 \\
\hline
\end{tabular}
\caption{\label{tab_countries} Foursquare data set by country. The first 15 countries sorted by the number of points of interest (POI) are shown, as well as the number of check-ins, users, and the percentage of the total dataset that represents. }
\end{table}
\begin{table}[t]
\centering
\begin{tabular}{|l|c|c|c|c|c|c|}
\hline
\multicolumn{1}{|c|}{ City}  & Country &    POIs &  Check-ins &  POIs / $\textrm{km}^2$ &  Check-ins PC \\
\hline
                   Istanbul &      TR &  334517 &    7343552 &      249.640 &      0.520404 \\
                    Jakarta &      ID &  398154 &    2908026 &       79.488 &      0.080083 \\
               Kuala Lumpur &      MY &  200199 &    2558716 &      150.752 &      0.403605 \\
                      Tokyo &      JP &  191162 &    2405876 &       35.946 &      0.072842 \\
                Mexico City &      MX &  137655 &    1804660 &       65.116 &      0.092265 \\
                    Bangkok &      TH &  169220 &    1741638 &       65.896 &      0.118231 \\
                     Moscow &      RU &  160868 &    1643199 &       85.477 &      0.116726 \\
                  São Paulo &      BR &  137494 &    1478616 &       68.576 &      0.077356 \\
                      Izmir &      TR &   74237 &    1443711 &      210.303 &      0.516692 \\
       Quezon City [Manila] &      PH &  123941 &    1343955 &       61.055 &      0.061959 \\
                   New York &      US &  137841 &    1311179 &       25.602 &      0.082202 \\
                   Santiago &      CL &   89115 &    1276748 &      123.771 &      0.201507 \\
                     Ankara &      TR &   59610 &    1021287 &      158.537 &      0.340152 \\
                    Bandung &      ID &  119357 &     933686 &      117.709 &      0.114121 \\
                  Singapore &      SG &   72981 &     827715 &       83.027 &      0.119596 \\
                   Surabaya &      ID &  110852 &     756628 &       63.308 &      0.090586 \\
                Los Angeles &      US &   91011 &     698605 &       16.157 &      0.048916 \\
              Osaka [Kyoto] &      JP &   66574 &     667631 &       21.081 &      0.042544 \\
                 Yogyakarta &      ID &   83635 &     609749 &       53.785 &      0.119592 \\
             Rio de Janeiro &      BR &   55765 &     570441 &       40.794 &      0.058220 \\
                      Belém &      BR &   39125 &     490162 &      143.842 &      0.234762 \\
                    Chicago &      US &   55319 &     485198 &       14.444 &      0.071565 \\
           Saint Petersburg &      RU &   56800 &     482715 &      107.780 &      0.112237 \\
                Kuwait City &      KW &   43547 &     473904 &       91.485 &      0.149590 \\
                     London &      GB &   48778 &     430524 &       26.168 &      0.044801 \\
                       Lima &      PE &   34989 &     404408 &       39.942 &      0.043646 \\
                   Denpasar &      ID &   54399 &     393148 &      130.453 &      0.209318 \\
                      Bursa &      TR &   25214 &     392454 &      120.067 &      0.233577 \\
                     Manaus &      BR &   34633 &     390084 &      132.693 &      0.193399 \\
                      Medan &      ID &   46528 &     387255 &       62.961 &      0.097886 \\
                      Seoul &      KR &   79306 &     382646 &       32.383 &      0.017714 \\
\hline
\end{tabular}
\caption{\label{tab_cities} Foursquare data by city. The 31 cities with the highest number of check-ins are shown. This selection contains information on a wide variety of countries with different geographic, cultural, linguistic, economic, and religious characteristics.}
\end{table}
Check-ins information was collected by Yang, et al. \cite{yang2020location, yang2020location2, yang2020lbsn2vec++, yang2019revisiting} and it is publicly available \cite{Yang-HomePage}. The datasets are divided into two sets. The first one is a list of check-ins obtained through an automated search on Twitter with the help of its API streaming service \cite{yang2020location}. Foursquare allows linking users’ accounts with other social media such as Twitter or Facebook, to share the check-ins on these platforms too. If this is the case, when one user registers their presence in a place, automatically, a tweet or a post is generated with the information of the check-in and some elements like hashtags and the URL of the venue’s page at Foursquare. Although the users that link their accounts in this way are only a subset of all the Foursquare users, this dataset contains information on about 2,733,324 users made in almost 22 months (from April 4, 2012, to January 29, 2014). With the information of tweets, a check-in dataset $D_{4S}$ was made with 90,048,627 rows, each one  corresponding to a check-in, i.e. the interaction between a user, with an anonymized ID, and a POI, where the ID is an alphanumeric identifier of the site in the Foursquare platform. Each record includes the Coordinated Universal Time (UTC) when the check-in occurred and the fourth column is the correction of the UTC corresponding to the Time Zone where the check-in was made. The user ID data has 2,733,324 different values, which corresponds to the activity of the same number of users. The POI ID column has 11,180,160 different values, which is equivalent to the number of POIs in the dataset.
\\[2mm]
From the information in the tweets, it is possible to obtain the POI profile from Foursquare’s site and with it the information contained in the second dataset, $D_{POI}$. Each POI ID is the same alphanumeric code that appears in $D_{4S}$; this column contains 11,180,160 different values. In addition, each place is described by its geographical coordinates: latitude, and longitude, described by floating numbers that obey the World Geodetic System 1984 standard (WGS84) with 6 decimal places, which is equivalent to a precision of less than one meter. Each POI is described by a category name; the dataset contains 519 different categories of which ``Home (private)'' is the most common with 1,310,012 sites, followed by ``Residential Building (Apartment / Condo)'' with 354,858; the third most popular is ``Office'' with 317,149; the fourth place is occupied by ``Building'' with 255,121 sites; the following are ``Café'', ``Restaurant'', ``Bar'' and ``Hotel'' with 188436, 153027, 145878 and 138476 sites, respectively. This information also includes the country code of the POI according to the two-letter ISO 3166-1 standard; containing 253 different codes. Then, the dataset includes check-in on every country in the world. From the 253 country codes, 84 countries with 5000 POIs or more represent 98.92\% of the data. Using the information in the country code, we group all the POIs by their code, obtaining 253 datasets, $D_{POI}(\textrm{code})$, each containing the POIs of a single country. In Table \ref{tab_countries}, 15 countries with most of the POIs are listed; 80\% of the venues belong to these countries. From this, we can highlight the presence of countries with remarkable diversity in terms of culture and geography.
\\[2mm]
With the information of each set $D_{POI}(\textrm{code})$, $D_{4S}$ can be filtered, grouping the check-ins by country code. In this manner, it is possible to know the number of check-ins per country and the number of users who had activity in each country, as shown in the Table \ref{tab_countries}. In addition to the number of POIs and check-ins, this table shows the percentage that this number represents of the total. The same is not done in the case of users since many users have activity in more than one country. Although the order in the ranking varies, the same countries that concentrate the majority of the POIs add up to the largest number of check-ins. Regarding check-ins, the records in the 85 countries that contain 25,000 check-ins or more, represent 99.54\% of the data.

\subsection*{Grouping POIs  by urban area}

Our topic of interest is the behavior of people in cities, so classifying POIs and check-ins by country is not enough. The definition of what a city is and what its borders are is a complex topic and has been dealt with by different authors \cite{bettencourt2021introduction, makse1, ortman2020cities}. In this work, we opt for the definition of {\it Functional Urban Area} used by the European Commission, which integrates factors such as infrastructure, population, and economy, and with which the Joint Research Centre generated the Global Human Settlement – Urban Centre Database (GHS), a dataset with the borders of 13,135 urban areas worldwide. This information is contained in a shapefile publicly available \cite{florczyk2019ghs} that includes the name of the city, its population, coordinates, area, the country, and region in which it is located, if it extends beyond the borders of a single region within the same country (New York, whose functional urban area is divided into counties belonging to the states of New York and New Jersey, in the United States) or even in more than one country (for example, Detroit, in the United States, whose functional urban area extends to Ontario, Canada, including the city of Windsor); among much more information. We use this dataset to group POIs by functional urban areas using Geopandas, a Python package for data analysis with geographic information \cite{GeoPandas}. Of all the urban areas contained in the GHS, 6,463 cities have, at least, one Foursquare POI. The POIs that are located within these cities represent 74 percent of the total; 82\% of the total check-ins were carried out in them. We focused on the 632 urban areas that have more than 10,000 check-ins, which represent 63\% of the total POIs (7,026,688), and 76\% of the total check-ins (68,356,896). That is, more than three-quarters of the total check-ins in the database were made in urban areas with more than 10,000 check-ins. The 31 cities with more records are shown in Table \ref{tab_cities}. Again, we find great diversity in cultural, social, and geographic terms, giving the Foursquare dataset great relevance for the study of urban dynamics. Along with the number of check-ins, the number of POIs per square kilometer and the number of check-ins per city inhabitant were calculated to give us an idea of the urban environments reflected by the Foursquare activity.
{
\subsection*{Geographic analysis of distances}
Each city, as well as each POI, has associated coordinates. To measure the distance between cities, the Haversine formula is required\cite{korn2000mathematical}. This formula calculates the physical distance between cities based on the great-circle distance between two points on a sphere, specifically the Earth's surface, as follows
\begin{equation}
	d = 2r \arcsin \left(\sqrt{  \sin^2\left( \frac{\varphi_2 - \varphi_1}{2}\right) + \cos \varphi_1\cdot \cos \varphi_2\cdot \sin^2\left( \frac{\lambda_2 - \lambda_1}{2} \right)  } \right),
\end{equation} where $\varphi_i$ and $\lambda_i$ are, respectively, the latitude and the longitude of the point $i$ and $r$ the radius of the sphere. To perform this calculation, we utilized the Haversine 2.8.0 library for Python\cite{HaversinePy}.
\subsection*{Kulback-Liebler divergence}
The Kullback-Leibler divergence, also known as relative entropy,  is an important quantity to calculate the difference between two probability distributions $P(z)$ and $Q(z)$ describing a stochastic variable $z$. For discrete distributions, this divergence is given by \cite{KullbackLiebler_1951,CoverThomasBook}
\begin{equation}\label{DKL_C}
	\mathcal{D}_{\mathrm{KL}}(P||Q) =  \sum_{z} P(z)\log\left[\frac{P(z)}{Q(z)}\right].
\end{equation}
Here $Q$ acts as a reference distribution. Also, it is important to emphasize that $\mathcal{D}_{\mathrm{KL}}(P||Q) $ is not a distance in the sense of a metric since the distance between $ P $ and $Q$ is not necessarily the same as between $Q$ and $P$. Also, from the definition in Eq. (\ref{DKL_C}), it is clear that $\mathcal{D}_{\mathrm{KL}}(P||Q)>0$ and is null when $P=Q$.
\subsection*{Networks and community detection}
Undirected networks with $N$ nodes are described by an adjacency $N\times N$ matrix $\mathbf{A}$ with entries $1$ if two different nodes are connected and $0$ otherwise. An important quantity in the study of networks is the degree of node $i$ given by  $k_{i}=\sum_{l=1}^N A_{il}$, which gives the number of connections to that node.
\\[2mm]
In many real networks, it is common to have subsets of nodes called communities. A community is defined as a locally dense connected subgraph in a network. The Louvain method is an algorithm to detect communities from large networks. The method optimizes the modularity $Q$, given by \cite{GirvanNewmanPRE2004,NewmanPNAS2006,Blondel_2008}
\begin{equation}
	Q = \frac{1}{2\mathcal{E}}\sum\limits_{i,j=1}^N\bigg[A_{ij} - \frac{k_i k_j}{2\mathcal{E}}\bigg]\delta_{c_i,c_j},
\end{equation}
where $c_i$, $c_j$ are the communities of the nodes $i$ and $j$, $\delta_{x,y}$ denotes the Kronecker delta and, $\mathcal{E}=\frac{1}{2}\sum_{i,j=1}^NA_{ij}$ is the total number of edges in the network.  
\\[2mm]
In the implementation of the method, the main goal is to generate a partition of the set of nodes in communities with labels $c_i$ for $i=1,2,\ldots,N$. The method works with the iteration of two steps: in the first step, for each node $i$, the change in modularity $\Delta Q$ is calculated for removing $i$ from its community and moving it into the community of each neighbor. Then, once this value is calculated for all communities, $i$ is placed into the community that resulted in the greatest modularity increase. If no increase is possible, the node $i$ remains in its original community. This process is applied repeatedly and sequentially to all nodes until no modularity increase can occur. Once a local maximum of modularity is reached, the first step is ended. In the second step, all the nodes in each community are grouped building a new network where nodes are the communities from the previous step. Any links between nodes of the same community are now represented by self-loops on the new community node and links from multiple nodes in the same community to a node in a different community are represented by weighted edges between communities. Once the new network is created, the second step has ended and the first step is applied to the new network. The values of $Q$ define a scale that measures the relative density of edges inside communities in comparison with the edges outside communities (see Ref. \cite{Fortunato2010Community,Blondel_2008} for details).
\subsection*{Machine learning and agglomerative clustering}
Machine learning (ML) is a branch of artificial intelligence that focuses on creating methods that can learn and improve their performance on tasks by leveraging data\cite{mehta2019high}. ML algorithms build models based on training data to make predictions or decisions. There are two main categories of Machine Learning algorithms: supervised and unsupervised.
In supervised learning, the algorithm is provided with labeled data, which consists of a set of input-output pairs. The goal of the algorithm is to learn a general rule or function that maps inputs into outputs. This is done by building a mathematical model of the data. The algorithm iteratively optimizes an objective function to learn a function that can accurately predict the output associated with new inputs.
\\[2mm]
In unsupervised learning, the algorithm is provided with unlabeled data and must identify patterns or structure in the data on its own\cite{mehta2019high}. Unsupervised learning algorithms discover patterns in the data and adapt their behavior based on the presence or absence of these patterns in new data. As a part of unsupervised learning, cluster analysis is the process of dividing a group of observations into subsets or clusters, where each one contains similar observations\cite{Agglomerative1}. The objective of clustering is to categorize untagged data into clusters based on a specific metric of similarity or distance. Essentially, a cluster is regarded as a collection of data points that exhibit some common pattern or structure. Clustering techniques vary in their assumptions on the data's structure and use different similarity metrics to evaluate the internal compactness and separation of clusters.
\\[2mm]
Hierarchical clustering is one of the most commonly used methods for grouping unlabeled data into clusters based on some similarity measure. This approach involves using either agglomerative or divisive algorithms to create nested clusters by either combining or separating previous clusters. Agglomerative clustering initially considers each data point as a separate cluster and merges them pairwise until all data is contained in a single cluster, or until a specific condition is met, such as grouping data into a specific number of clusters. To accomplish this, a distance metric is used to determine the distance between each pair of data points, and a criterion for determining which clusters to merge at each stage is employed. In this study, the Euclidean distance was used to measure the distance between check-in distributions, and the criterion for merging clusters was to combine the two whose union minimized the variance of distances within all clusters. To do this, the Scikit-learn\cite{scikit-learn} python library was used.
}

\section*{Acknowledgements}

 F.B. acknowledges support from CONACYT M\'exico. This work was supported by PAPIIT-UNAM grant No. IN116220.

\section*{Author contributions statement}

F.B., A.P.R. and J.L.M. designed the research, performed the research, and wrote the manuscript.

\section*{Data availability}
The datasets analysed during the current study are available in the webpage: \url{https://sites.google.com/site/yangdingqi/home/foursquare-dataset}

\end{document}